\def\final{0}

\documentclass[11pt]{article}

\usepackage{color,fullpage}
\usepackage{paralist}
\usepackage{times}
\usepackage{amsmath,amsfonts,amsthm,graphicx}
\usepackage[numbers]{natbib}
\usepackage{hyperref}

\ifnum\final=0
\newcommand{\mnote}[2]{{\textcolor{#1}{ #2}}}
\else
\newcommand{\mnote}[2]{}
\fi

\ifnum\final=0
\newcommand{\mynote}[1]{\marginpar{\tiny\sf #1}}
\else
\newcommand{\mynote}[1]{}
\fi

\newcommand{\sampnum}{\mathcal{A}}
\newcommand{\AAA}{\mathcal{A}}

\newcommand{\remove}[1]{}
\newcommand{\ignore}[1]{}

\DeclareMathSymbol{\R}{\mathbin}{AMSb}{"52}

\newcommand{\set}[1]{\left\{ {#1} \right\}}
\newcommand{\paren}[1]{\left( {#1} \right)}

\newtheorem{theorem}{Theorem}[section]

\newtheorem{lemma}[theorem]{Lemma}

\newtheorem{corollary}[theorem]{Corollary}

\newtheorem{myexample}{Example}
\newenvironment{example}{\begin{myexample}\sf\rm}{$\diamondsuit$\end{myexample}}
\newtheorem{myremark}{Remark}
\newenvironment{remark}{\begin{myremark}\sf\rm}{$\diamondsuit$\end{myremark}}

\newtheorem{definition}[theorem]{Definition}

\newcommand{\thmref}[1]{Theorem~\ref{thm:#1}}

\newcommand{\lemref}[1]{Lemma~\ref{lem:#1}}

\newcommand{\remref}[1]{Remark~\ref{rem:#1}}
\newcommand{\defref}[1]{Definition~\ref{def:#1}}

\newcommand{\secref}[1]{Section~\ref{sec:#1}}

\newcommand{\appref}[1]{Appendix~\ref{app:#1}}

\newcommand{\eps}{\epsilon}

\newcommand{\logdel}{\log\paren{\tfrac{1}{\delta}}}

\newcommand{\half}{\frac{1}{2}}

\DeclareMathSymbol{\erert}{\mathbin}{AMSb}{"50}

\newcommand{\A}{{\cal A}}

\newcommand{\E}{{\mathbb{E}}}

\newcommand{\sd}[1]{\mathbf{SD}\paren{{#1}}}

\newcommand{\beq}{\begin{equation}}
\newcommand{\eeq}{\end{equation}}
\newcommand{\bml}{{\begin{multline}}}
\newcommand{\eml}{{\end{multline}}}
\DeclareMathSymbol{\Q}{\mathbin}{AMSb}{"51}

\newcommand{\D}{{\mathcal{D}}}
\newcommand{\dx}{{\mathrm{x}}}
\newcommand{\dz}{{\mathrm{z}}}
\newcommand{\dy}{{\mathrm{y}}}

\newcommand{\dham}{{d_H}}

\newenvironment{CompactEnumerate}{\begin{enumerate}}{\end{enumerate}}

\begin{document}
\title{On the `Semantics' of Differential Privacy: \\
A Bayesian Formulation\thanks{Preliminary statements of the main results from this paper appeared in \citet{GKS08}. This paper contains strengthened results as well as full proofs.}}

\author{Shiva Prasad Kasiviswanathan\thanks{S.K.'s work at Penn State was partly
supported by NSF award TF-0729171. S.K. is now at Amazon Research.}\\ General Electric Global Research\\ {\tt kasivisw@gmail.com} \and Adam Smith\thanks{A.S. was supported by NSF awards
TF-0729171, CDI-0941553, PECASE CCF-0747294 and a Google Faculty
Award. A.S. is now at Boston University.} \\ Department of Computer Science and Engineering \\
The Pennsylvania State University\\
{\tt ads22@bu.edu}}
\date{}
\maketitle
\begin{abstract}
Differential privacy is a definition of ``privacy'' for algorithms that analyze and publish information about statistical databases. It is often claimed that differential privacy provides guarantees against adversaries with arbitrary side information. In this paper, we provide a precise formulation of these guarantees in terms of the inferences drawn by a Bayesian adversary. We show that this formulation is satisfied by both $\eps$-differential privacy as well as a relaxation, $(\eps,\delta)$-differential privacy. Our formulation follows the ideas originally due to Dwork and McSherry, stated implicitly in \cite{Dwork06}. This paper is, to our knowledge, the first place such a formulation appears explicitly. The analysis of the relaxed definition is new to this paper, and provides guidance for setting the $\delta$ parameter when using  $(\eps,\delta)$-differential privacy. 
\end{abstract}


\section{Introduction}
Privacy is an increasingly important aspect of data publishing. Reasoning about privacy, however, is fraught with pitfalls. One of the most significant is the auxiliary information (also called external knowledge, background knowledge, or side information) that an adversary gleans from other channels such as the web, public records, or domain knowledge.  Schemes that retain privacy guarantees in the presence of independent releases are said to {\em compose securely}. The terminology, borrowed from cryptography (which borrowed, in turn, from software engineering), stems from the fact that schemes that compose securely can be designed in a stand-alone fashion without explicitly taking other releases into account. Thus, understanding independent releases is essential for enabling modular design. In fact, one would like schemes that compose securely not only with independent instances of themselves, but with {\em arbitrary external knowledge}.

Certain randomization-based notions of privacy (such as \emph{differential
privacy}, due to~\citet*{DMNS06}) are viewed as providing meaningful guarantees
even in the presence of arbitrary side information. In this paper. we
give a precise formulation of this statement. First, we provide a
Bayesian formulation of ``pure'' differential privacy which explicitly
models side information. Second, we prove that the relaxed definitions
of \citet{BDMN05,DKMMN06} and \citet{MKAGV08} imply the Bayesian formulation.  The
proof is non-trivial, and relies on the ``continuity'' of Bayes' rule
with respect to certain distance measures on probability
distributions. Our result means that techniques satisfying the
relaxed definitions can be used with the same sort of assurances as in
the case of pure differentially-private algorithms, as long parameters are set appropriately. 
Specifically, 
$(\eps,\delta)$-differential privacy provides meaningful guarantees
whenever $\delta$, the additive error parameter, is smaller than
about $\eps^2/n$, where $n$ is the size of the data set.

\paragraph{Organization.}
After introducing the basic definitions, we state and discuss our
main results in \secref{bayes}. In \secref{related}, we relate our
approach to other
efforts---subsequent to the initial version of this work---that sought
to pin down mathematical precise formulations of the 
``meaning'' of differential privacy.
\secref{bigproofs} proves our main theorems. Along the way, we develop
lemmas about \emph{$(\eps,\delta)$-indistinguishability}---the notion
of similarity that underlies $(\eps,\delta)$-differential
privacy----that we believe are of independent interest. The most
useful of these, which we dub the Conditioning Lemma, is given in \secref{conditioning}. 
Finally, we provide
further discussion of our approach in \secref{discussion}. 

\subsection{Differential Privacy}
Databases are assumed to be vectors in $\mathcal{D}^n$ for some domain $\mathcal{D}$. The Hamming distance $\dham(\dx,\dy)$ on $\mathcal{D}^n$ is the number of positions in which the vectors $\dx,\dy$ differ. We let $\Pr[\cdot]$ and $\E[\cdot]$ denote probability and expectation, respectively. Given a randomized algorithm $\mathcal{A}$, we let $\mathcal{A}(\dx)$ be the random variable (or, probability distribution on outputs) corresponding to input $\dx$.  If $X$ and $Y$ are probability distributions (or random variables) on a discrete space $D$, the {\em statistical difference} (a.k.a.\ {\em total variation distance})  between $X$ and $Y$ is defined as:
$$\sd{X,Y}= \max_{S \subset D}\left |\Pr[X \in S]-\Pr[Y \in S] \right |.$$

\begin{definition}[$\eps$-differential privacy \citep{DMNS06}] \label{def:ind}
A randomized algorithm $\sampnum$ is said to be $\eps$-differentially private if for all databases $\dx,\dy\in \mathcal{D}^n$ at Hamming distance at most 1, and for all subsets $S$ of outputs,
$$
\Pr[\sampnum (\dx)\in S] \leq e^{\eps} \cdot \Pr[\sampnum (\dy)\in S].
$$
\end{definition}

This definition states that changing a single individual's data in the database leads to a small change in the {\em distribution} on outputs. Unlike more standard measures of distance such as statistical difference or Kullback-Leibler divergence, the metric here is
multiplicative and so even very unlikely events must have approximately the same probability under the distributions $\AAA(\dx)$ and $\AAA(\dy)$. This condition  was relaxed somewhat in other papers \citep{DiNi03,DwNi04,BDMN05,DKMMN06,CM06,NRS07,MKAGV08}. The schemes in all those papers, however,  satisfy the following relaxation~(first formulated by \citet*{DKMMN06}):

\begin{definition}[$(\eps,\delta)$-differential privacy \citep{DKMMN06}] 
\label{def:indd}
A randomized algorithm $\sampnum$ is $(\eps,\delta)$-differentially
private if for all databases $\dx,\dy\in \D^n$ that differ in one
entry, and for all subsets $S$ of outputs, $$\Pr[\sampnum (\dx)\in S]
\leq e^{\eps} \cdot \Pr[\sampnum (\dy)\in S]+\delta\,.$$
\end{definition}

\section{Semantics of Differential Privacy} \label{sec:bayes}
There is a crisp, semantically-flavored\footnote{The use of the term
  ``semantic'' for definitions that deal directly with adversarial
  knowledge dates back to \emph{semantic security} of encryption
  \citep{GM84}.} interpretation of differential privacy, due to Dwork
and McSherry, explained in \cite{Dwork06}: {\em Regardless of external
  knowledge, an adversary with access to the sanitized database draws
  the same conclusions whether or not my data is included in the
  original database.} One might hope for a stronger statement, namely
that the adversary draws the same conclusions whether or not the data
is used at all. However, such a strong statement is impossible to
provide in the presence of arbitrary external information
(\citet{DN08-imposs,Dwork06}; see also \citet{KiferM11}), as
illustrated by the following example.

\begin{example}\label{ex:smoking}
  Consider a clinical study that explores the
  relationship between smoking and lung disease. A health insurance
  company who had no a priori understanding of that relationship might
  dramatically alter its ``beliefs'' (as encoded by insurance
  premiums) to account for the results of the study. The study would
  cause the company to raise premiums for smokers and lower them for
  nonsmokers, regardless of whether they participated in the study. In
  this case, the conclusions drawn by the company about the riskiness
  of any one individual (say Alice) are strongly affected by the
  results of the study.
  This occurs regardless of whether Alice's data are included in the study.
\end{example}

In this section, we develop a formalization Dwork and McSherry's
interpretation and explore its relation to standard definitions. 
To proceed, we require a mathematical formulation of ``external knowledge'', and of ``drawing conclusions''. The first is captured via a {\em prior} probability distribution $b$ on $\D^n$ ($b$ is a mnemonic for ``beliefs''). Conclusions are modeled by the corresponding {\em posterior} distribution: given a transcript $t$, the adversary updates his belief $b$ about the database $\dx$ using Bayes' rule to obtain a posterior $\bar{b}$:
\begin{eqnarray} \label{eqn:bel}
\bar{b}[\dx| t] = \frac{\Pr[\sampnum (\dx)=t] b[\dx]}{\sum_\dz \Pr[\sampnum (\dz)=t]b[\dz]}\ . \end{eqnarray}

When the mechanism $\A$ is interactive, the definition of $\sampnum$
depends on the adversary's choices; for legibility we omit the
dependence on the adversary in the notation. Also, for simplicity, we
discuss only discrete probability distributions. Our results extend
directly to the interactive, continuous case.

For a database $\dx$, define $\dx_{-i}$ to be the same vector except
position $i$ has been replaced by some fixed, default value $\D$. Any
valid value in $\D$ will do for the default value. We define $n+1$
related games, numbered 0 through $n$. In Game 0, the adversary
interacts with $\sampnum(\dx)$. This is the interaction that actually
takes place between the adversary and the randomized algorithm
$\AAA$. The distribution $\bar{b}_0$ is just the distribution
$\bar{b}$ as defined in~\eqref{eqn:bel}.

In Game $i$ (for $1\leq i \leq n$), the adversary interacts with
$\sampnum(\dx_{-i})$. Game $i$ describes the hypothetical scenario
where person $i$'s data is not used.\!\footnote{It could happen by
  coincidence that person $i$'s data equals the default value and
  hence that $\dx=\dx_{-i}$. This doesn't affect the meaning of the
  result since the default value is chosen independently of the
  data. Readers bothered by the possible coincidence may choose to
  think of the default value as a special value $\perp$ (e.g., ``no
  data'') that does not correspond to any real record.} In Game $i > 0$, given a transcript $t$, the adversary updates his belief $b$ about database $\dx$ again using Bayes' rule to obtain a posterior~$\bar{b}_i$ as follows:
\begin{eqnarray} \label{eqn:beli}
\bar{b}_i[\dx| t] = \frac{\Pr[\sampnum (\dx_{-i})=t] b[\dx]}{\sum_\dz \Pr[\sampnum (\dz_{-i})=t]b[\dz]}. 
\end{eqnarray}

Through these $n+1$ games, we get $n+1$ {\sl a posteriori} distributions $\bar{b}_0,\dots,\bar{b}_n$, where $\bar{b}_0$ is same as $\bar{b}$ (defined in \eqref{eqn:bel}), and $\bar{b}_i$ ($i > 0$) is the posterior distribution obtained when the adversary interacts with $\A(\dx_{-i})$ and uses this interaction to update his belief distribution (defined in~\eqref{eqn:beli}).

Given a particular transcript $t$, we say privacy has been breached if
the adversary would draw different conclusions about the world and, in
particular, about a person $i$, depending on whether or not $i$'s data
was used. One could formally define ``different'' in many ways. In
this paper, we choose a weak (but popular) measure of distance between
probability distributions, namely statistical difference. We say the
adversary has learned something, if for any transcript $t$ the distributions $\bar{b}_0[\cdot|t]$ and $\bar{b}_i[\cdot|t]$ are far apart in statistical difference. We would like to avoid this from happening for any potential participant. This is captured by the following definition.

\begin{definition}[$\eps$-semantic privacy]  \label{def:sem}
A randomized algorithm $\sampnum$ is said to be $\eps$-semantically private if for all belief distributions $b$ on $\mathcal{D}^n$, for all possible transcripts $t$, and for all $i = 1,\dots,n$$\mathrm{:}$
$$\sd{\bar{b}_0[\cdot|t]\ ,\  \bar{b}_i[\cdot|t]\ } \leq \eps.$$
\end{definition}


Our formulation of semantic privacy is inspired by Dwork and
McSherry's interpretation of differential privacy~\citep{Dwork06}. We now formally show that the notions of $\eps$-differential privacy (Definition \ref{def:ind}) and $\eps$-semantic privacy (Definition \ref{def:sem})
are essentially equivalent.

\begin{theorem}\label{thm:eind} For all $\eps>0$,
$\eps$-differential privacy implies $\bar\eps$-semantic privacy,
where $\bar\eps=e^{\eps}-1$. For $0<\eps\leq0.45$,  $\eps/2$-semantic privacy implies $3\eps$-differential privacy.
\end{theorem}

The proof of this and all other results in this section may be found
in \secref{bigproofs}.

We can extend the previous Bayesian formulation to capture situations
where bad events can occur with some negligible
probability. Specifically, we formulate $(\eps,\delta)$-semantic privacy and show that it is closely related to $(\eps,\delta)$-differential privacy.

\begin{definition}[($\eps,\delta$)-semantic privacy] \label{defn:epsdelta}
A randomized algorithm is $(\eps,\delta)$-semantically private if for
all belief distributions $b$ on $\mathcal{D}^n$,  with probability at
least $1-\delta$ over $t\sim\AAA(\dx)$ ($t$ drawn from $\AAA(\dx)$),  where the database $\dx$
is drawn according to $b$, and for all $i =1,\dots,n$:
$$\sd{\bar{b}_0[\cdot|t]\ ,\  \bar{b}_i[\cdot|t]\ } \leq \eps. $$
\end{definition}

The $(\eps,\delta)$-privacy definition is most interesting when
$\eps\gg \delta$, since every $(\eps,\delta)$-private algorithm is
also $(0, \delta + (e^\eps-1))$-differentially private. Below, we
assume $\eps>\delta$. 
In fact, many of our results are meaningful only when $\delta$ is less
than $1/n$, while $\eps$ must generally be much larger than $1/n$ to
allow for useful algorithms.

\begin{theorem}[Main Theorem]\label{thm:ind2sdp} {~}
  \renewcommand{\labelenumi}{(\arabic{enumi})} 
  \begin{enumerate}
\item 
    If $\eps,\delta>0$ and $\delta< (1- e^{-\eps})^2/n$, then
    ($\eps,\delta$)-differential privacy implies
    $(\eps',\delta')$-semantic privacy on databases of size $n$ with
    $\eps'=e^{3\eps}-1+2\sqrt{n\delta}$ and $\delta' =
    4\sqrt{n\delta}$. 
\item If $\eps,\delta>0$ and $\eps \leq 0.45$, then
    $(\eps,\delta)$-semantic privacy implies
    $(3\eps,2\delta)$-differential privacy.
  \end{enumerate}
\end{theorem}

In~\appref{anotherview}, we discuss a stronger notion of $(\eps,\delta)$-semantic privacy and show that $(\eps,\delta)$-differential privacy need not imply this stronger semantic privacy guarantee.

\begin{remark}\label{rem:statdiff}
  The implications in Theorems \ref{thm:eind} and \ref{thm:ind2sdp}
  would not hold if differential
  privacy were defined in terms of statistical difference (total
  variation distance) or mutual information instead of the
  multiplicative metric used in Definitions~\ref{def:ind}
  and~\ref{def:indd}. For example, one could change the last line of
  the Definition~\ref{def:indd} to
  \begin{equation}
 \Pr[\sampnum (\dx)\in S] \leq  \Pr[\sampnum (\dy)\in
 S]+\eps_{SD}\,.\label{eq:1}
\end{equation}
For this modified definition to allow publishing useful information, one would need $\eps_{SD}
=\Omega( 1/n)$ (otherwise, data sets that differ in all $n$ elements
would still be hard to distinguish). However, in that parameter range
there is a mechanism that satisfies the new definition but does not
satisfy ``semantic'' privacy for any reasonable parameters. Namely,
consider the mechanism which on input $\dx=(x_1,\dots,x_n)$ samples a uniformly random index $i\in
\{1,...,n\}$ and outputs $(i,x_i)$. This mechanism is intuitively
unsatisfactory, since it always outputs some
individual's data in the clear. It also does not satisfy semantic privacy for any pair $(\eps,\delta)$
where $\eps<1$ and $\delta<1$. Nevertheless, it does satisfy the
requirement of \eqref{eq:1} with $\epsilon_{SD} =1/n$. 
The same mechanism also satisfies the natural variant of differential
privacy based on mutual information (for example, where the mutual information
between $\sampnum(\dx)$ and $x_i$ is required to be small for all indices
$i$ and product distributions on $\dx$). 
\end{remark}

\subsection{Related Approaches}
\label{sec:related}

\paragraph{Prior to Posterior Comparisons.}
In the original paper on differential privacy, \citet{DMNS06} defined
a notion of ``semantic'' privacy that involved comparing the prior and
posterior distributions of the adversary. In the language of the
preceding section, they require that $\sd{b[\cdot]\ ,\
  \bar{b}_i[\cdot|t]\ } \leq \eps$ for a subclass of belief
distributions, called ``informed beliefs'', in which all but one of the
data set entries are fixed (constant).  They show that this definition
is equivalent to differential privacy. 
\citet{KiferM12} use this prior-to-posterior approach to generalize
differential privacy to other settings.

However, the impossibility results of Dwork and Naor
\citep{Dwork06,DN10} and \citet{KiferM11}, exemplified by the smoking example in
Example~\ref{ex:smoking}, imply that no mechanism
that provides nontrivial information about the data set satisfies
such a prior-to-posterior definition for \emph{all}
distributions. 

This impossibility motivated the posterior-to-posterior comparison
espoused in this paper, and subsequently generalized by
\citet{BassilyGKS13}. In contrast to the prior-to-posterior approach, the
framework discussed in this paper does generalize to arbitrary
distributions on the data (and, hence, to arbitrary side
information). \citet{BassilyGKS13}~suggest the term ``inference-based'' for
definitions which explicitly discuss the posterior distributions
constructed by Bayesian adversaries.

\paragraph{Hypothesis Testing.} 
\citet{WZ10} relate differential privacy to the type I and II errors
of a hypothesis test. Specifically, fix an 
$\eps$-differentially private mechanism $\sampnum$, an i.i.d.
distribution on the data $\dx$, an index $i$, and disjoint sets $S$
and $T$ of possible
values for the $i$-th entry $x_i$ of $\dx$. \citet{WZ10} show that any
hypothesis test (given $\sampnum(\dx)$, and full knowledge of the input product distribution on $\dx$ and the differentially private mechanism $\sampnum$) for the hypothesis $H_1: x_i\in S$ versus the alternative
$H_1: x_i \in T$ must satisfy 
\begin{equation}
1-\beta\leq e^{\eps}\alpha\, ,\label{eq:3}
\end{equation}
where $\alpha$ is the significance level (maximum type-I error) and $1-\beta$
is the power (maximum type-II error) of the test.
In other words, the test rejects the hypothesis with approximately the same
probability regardless of whether the hypothesis is true. This perspective was extended to $(\eps,\delta)$-differential privacy by~\citet{HRW13}.

This is a
reasonable requirement.  Note, however, that it holds only for
product distributions, which limits its applicability. More
importantly, a  very similar
statement can be proven for the statistical difference-based
definition discussed in \remref{statdiff}. Specifically, one can show that 
\begin{equation}
 1-\beta\leq 
\alpha
+\eps_{SD}\,.\label{eq:2}
\end{equation}
when the mechanism satisfies the definition of
\remref{statdiff}. Equation \eqref{eq:3} has the same natural language
interpretation as equation \eqref{eq:2}, namely, ``the test rejects the hypothesis with approximately the same
probability regardless of whether the hypothesis is true''. 
However, as mentioned in the \remref{statdiff}, the statistical difference-based
definition allows mechanisms that publish detailed personal data in the
clear. This makes the meaning of a hypothesis-testing-based
definition hard to evaluate intuitively. We hope the definitions
provided here
are easier to interpret.

\section{Proofs of Main Results}
\label{sec:bigproofs}

We begin this section by defining
\emph{$(\eps,\delta)$-indistinguishability} and stating a few of its
basic properties (Section~\ref{sec:basicprops}, with proofs in
\appref{propertiesproof}). \secref{pure-proof} gives the proof of our
main result for $\eps$-differential privacy. In
Section~\ref{sec:conditioning} we state and prove the Conditioning
Lemma, the main tool which allows us to prove our results about
$(\eps,\delta)$-differential privacy
(Section~\ref{sec:approxproof}). 

\subsection{$(\eps,\delta)$-Indistinguishability and its Basic Properties}
\label{sec:basicprops}

The relaxed notions of $(\eps,\delta)$-differential privacy implicitly uses a two-parameter distance measure on probability distributions (or random variables) which we call $(\eps,\delta)$-indistinguishability. In this section, we develop a few basic properties of this measure. These properties listed in Lemma~\ref{lem:properties} will play an important role in establishing the proofs of Theorems~\ref{thm:eind} and~\ref{thm:ind2sdp}

\begin{definition} [$(\eps,\delta)$-indistinguishability]
Two random variables $X,Y$ taking values in a set $D$ are $(\eps,\delta)$-indistinguishable if for all sets $S\subseteq D$,
\begin{eqnarray*}
\Pr[X\in S] \leq e^{\displaystyle \eps} \Pr[Y\in S] + \delta \ \ \ \ \mbox{ and } \ \ \ \ \Pr[Y\in S] \leq e^{\displaystyle \eps} \Pr[X\in S] + \delta.
\end{eqnarray*}
\end{definition}

We will also be using a variant of $(\eps,\delta)$-indistinguishability, which we call {\em point-wise} $(\eps,\delta)$-indistinguishability. \lemref{properties} (Parts~\ref{it:pw2ind} and \ref{it:ind2pw}) shows that $(\eps,\delta)$-indistinguishability and point-wise $(\eps,\delta)$-indistinguishability are almost equivalent. 
\begin{definition} [Point-wise $(\eps,\delta)$-indistinguishability] 
Two random variables $X$ and $Y$ are point-wise $(\eps,\delta)$-indistinguishable if with probability at least $1-\delta$ over $a$ drawn from either $X$ or $Y$, we have: \[e^{-\eps}\Pr[Y=a] \leq \Pr[X=a] \leq e^{\eps} \Pr[Y=a].\]
\end{definition}

\begin{lemma}\label{lem:properties}
Indistinguishability satisfies the following properties:
\begin{CompactEnumerate}
\item \label{it:pw2ind}
If $X,Y$ are point-wise $(\eps,\delta)$-indistinguishable then they are $(\eps,\delta)$-indistinguishable.

\item \label{it:ind2pw}
If $X,Y$ are $(\eps,\delta)$-indistinguishable then they are
point-wise $\left(2\eps\ ,\ \delta \cdot \frac{2}{1-e^{-\eps}}\right)$-indistinguishable.

\item \label{it:ajoint}
Let $X$ be a random variable on $D$. Suppose that for every $a \in D$, $\AAA(a)$ and $\AAA'(a)$ are $(\eps,\delta)$-indistinguishable (for some randomized algorithms $\AAA$ and $\AAA'$). Then the pairs $(X,\AAA(X))$ and $(X,\AAA'(X))$ are $(\eps,\delta)$-indistinguishable. 

\item \label{it:joint}
Let $X$ be a random variable.  Suppose with probability at least $1-\delta_1$ over $a \sim X$, $\sampnum (a)$ and $\sampnum'(a)$ are $(\eps,\delta)$-indistinguishable (for some randomized algorithms $\AAA$ and $\AAA'$). Then the pairs $(X,\ \sampnum (X))$ and $(X,\ \sampnum '(X))$ are $(\eps,\delta+\delta_1)$-indistinguishable.


\item \label{it:sd} If $X,Y$ are $(\eps,\delta)$-indistinguishable (or $X,Y$ are point-wise $(\eps,\delta)$-indistinguishable), then $\sd{X,Y} \leq \bar\eps+\delta$, where $\bar\eps=e^{\eps}-1$.

\end{CompactEnumerate}
\end{lemma}
The lemma is proved in \appref{propertiesproof}.


\subsection{Case of $\eps$-Differential Privacy: Proof of \thmref{eind}}
\label{sec:pure-proof}

\newtheorem*{t1}{Theorem~\ref{thm:eind} (restated)}
\begin{t1}
[Dwork-McSherry]  $\eps/2$-differential privacy implies $\bar\eps$-semantic privacy, where $\bar\eps=e^{\eps}-1$. $\eps/2$-semantic privacy implies $3\eps$-differential privacy as long as $\eps \leq 0.45$. 
\end{t1}
\begin{proof}
  Consider any database $\dx \in \mathcal{D}^n$. Let $\AAA$ be an
  $\eps/2$-differentially private algorithm. Consider any belief
  distribution $b$. Let the posterior distributions $\bar{b}_0[\dx|t]$
  and $\bar{b}_i[\dx|t]$ for some fixed $i$ and $t$ be as defined
  in~\eqref{eqn:bel} and~\eqref{eqn:beli}. $\eps/2$-differential
  privacy implies that for every database $\dz \in \mathcal{D}^n$
  \[ e^{-\eps/2} \Pr[\sampnum (\dz_{-i})=t] \leq \Pr[\sampnum(\dz)=t]
  \leq e^{\eps/2} \Pr[\sampnum (\dz_{-i})=t].\]
  These inequalities imply that the ratio of $\bar{b}_0[\dx|t]$ and
  $\bar{b}_i[\dx|t]$ (defined in \eqref{eqn:bel} and~\eqref{eqn:beli})
  is within $e^{\pm \eps}$.  Since these inequalities holds for every
  $\dx$, we get:
  \[ \forall \dx \in \mathcal{D}^n, \,\, e^{-\eps}\bar{b}_i[\dx|t]
  \leq \bar{b}_0[\dx|t] \leq e^{\eps}\bar{b}_i[\dx|t]. \] 
  This implies
  that the random variables (distributions) $\bar{b}_0[\cdot|t]$ and
  $\bar{b}_i[\cdot|t]$ are point-wise $(\eps,0)$-indistinguishable.
  Applying \lemref{properties} (Part~\ref{it:sd}) with $\delta = 0$,
  gives $\sd{\bar{b}_0[\cdot|t],\bar{b}_i[\cdot|t]} \leq
  \bar\eps$. Repeating the above arguments for every belief
  distribution, for every $i$, and for every $t$, shows that $\AAA$ is
  $\bar\eps$-semantically private.

To see that $\eps$-semantic privacy implies $3\eps$-differential privacy, consider a belief distribution $b$ which is uniform over two databases $\dx,\dy$ which are at Hamming distance of one. Let $i$ be the position in which $\dx$ and $\dy$ differ. Fix a transcript $t$. The distribution $\bar{b}_i[\cdot|t]$ will be uniform over $\dx$ and $\dy$ since they induce the same distribution on transcripts in Game $i$. This means that $\bar{b}_0[\cdot|t]$ will assign probabilities $1/2 \pm \eps$ to each of the two databases (by~\defref{sem}). Working through Bayes' rule shows that (note that $b[\dx]=b[\dy]$) 
\begin{eqnarray} \label{eqn:bineq}
\frac{\Pr[\sampnum (\dx)=t]}{\Pr[\sampnum (\dy)=t]}=
\frac{\bar{b}_0[\dx|t]}{\bar{b}_0[\dy|t]} \leq
\frac{\half+\eps}{\half-\eps} \leq e^{3\eps} \text{(since $\eps\leq 0.45$).}
\end{eqnarray} 
Since the bound in~\eqref{eqn:bineq} holds for every $t$, $\AAA(\dx)$ and $\AAA(\dy)$ are point-wise $(3\eps,0)$-indistinguishable. Using \lemref{properties} (Part~\ref{it:pw2ind}), implies that $\AAA(\dx)$ and $\AAA(\dy)$ are $(3\eps,0)$-indistinguishable. Since this relationship holds for every pair of neighboring databases $\dx$ and $\dy$, means that $\AAA$ is $3\eps$-differentially private.
\end{proof}

\subsection{A Useful Tool: The Conditioning Lemma} 
\label{sec:conditioning}

We will use the following lemma to establish connections between $(\eps,\delta)$-differential privacy and $(\eps,\delta)$-semantic privacy. Let $B|_{A=a}$ denote the conditional distribution of $B$ given that $A=a$ for jointly distributed random variables $A$ and $B$.
\begin{lemma}[Conditioning Lemma]\label{lem:bayes}
  Suppose the pair of random variables $(A,B)$ is
  $(\eps,\delta)$-indistinguishable from the pair $(A',B')$. Then, for
  $\hat\eps=3\eps$ and for every $\hat\delta>0$, the following holds:
  with probability at least $1-\delta''$ over $t\sim B$ (or,
  alternatively, over
  $t \sim B'$), the random variables $A|_{B=t}$ and $A'|_{B'=t}$ are
  $(\hat\eps,\hat\delta)$-indistinguishable, where
  $\delta''=\frac{2\delta}{\hat\delta}+\frac{2\delta}{1- e^{-\eps}}$.
\end{lemma}

We can satisfy the conditions of the preceding lemma by setting $\hat
\delta = \delta'' = O(\sqrt{\delta})$ for any constant $\eps$.  However, the proof of our main theorem will use a slightly different setting (with $\delta''$ smaller than $\hat\delta$).

\begin{proof} 
Let $(A,B)$ and $(A',B')$ take values in the set $D \times E$.  In the remainder of the proof, we will use the notation $A|_t$ for $A|_{B=t}$ and $A'|_t$ for $A'|_{B'=t}$.  
Define,
\begin{eqnarray*} 
&Bad_1=\set{t \in E \, : \, \exists S_t \subset D \mbox{ such that } \Pr[A|_t \in S_t] > e^{\hat\eps}\Pr[A'|_t \in S_t] + \hat\delta}& \\ 
&Bad_2=\set{t \in E\, : \, \exists S_t \subset D \mbox{ such that } \Pr[A'|_t \in S_t] > e^{\hat\eps}\Pr[A|_t \in S_t] + \hat\delta}.& \end{eqnarray*}
To prove the lemma, it suffices to show that the probabilities $\Pr[B \in Bad_1 \cup
Bad_2]$ and $\Pr[B' \in Bad_1 \cup Bad_2]$ are each at most
$\delta''$. To do so, we first consider the set  
$$Bad_0 = \set{t\in E\,:\, \Pr[B=t] <e^{-2\eps}\Pr[B'=t] \text{ or }\Pr[B=t] >e^{2\eps}\Pr[B'=t] }\,.$$
We will separately bound the probabilities of $Bad_0$, $Bad_1' =
Bad_1\setminus Bad_0$ and $Bad_2' = Bad_2 \setminus Bad_0$.

To bound the mass of $Bad_0$,  note that $B$ and $B'$ are
$(\eps,\delta)$-indistinguishable (since they are functions of $(A,B)$
and $(A',B')$)\footnote{\emph{Note:} Even if we started with the stronger
  assumption that the pairs $(A,B)$ and $(A',B')$ are point-wise
  indistinguishable, we would still have to make a nontrivial argument
  to bound $Bad_0$, since point-wise indistinguishability is \emph{not}, in general,
  closed under postprocessing.}. Since $(\eps,\delta)$-indistinguishability implies point-wise $(2\eps,\frac{2\delta}{1-e^{-\eps}})$-indistinguishability (\lemref{properties}, Part ~\ref{it:ind2pw}),
we have
\[ \Pr[B \in Bad_0] \leq \frac{2\delta}{1-e^{-\eps}}\,.\]
We now turn to $Bad_1'=Bad_1 \setminus Bad_0$.
For each $t \in Bad_1'$, let $S_t$ be any set that witnesses $t$'s
membership in $Bad_1$ (that is, for which $\Pr[A|_t\in S_t] $ exceeds
$e^{\hat\eps}\Pr[A'|_t \in S_t] + \hat\delta$). Consider the
critical set $$T_1 = \bigcup_{t \in Bad_1'}  (S_t \times \{t\})\,.$$ 
Intuitively, this set will have large mass if $Bad_1'$ does. 
Specifically, by the definition of $S_t$, we get a lower bound on the probability of $T_1$:
\begin{eqnarray*}
\Pr[(A,B)\in T_1] &=& \sum_{t \in Bad_1'} \Pr[A|_t \in S_t] \Pr[B=t]  \\
&>& \sum_{t \in Bad_1'} (e^{\hat\eps}\Pr[A'|_t \in S_t]+\hat\delta)\Pr[B=t] \\
&=&\paren{\sum_{t \in Bad_1'} e^{\hat\eps}\Pr[A'|_t \in S_t]\Pr[B=t]}
+ \hat\delta\Pr[B \in Bad_1']. 
\end{eqnarray*}
Because $Bad_1'$ does not contain points in $Bad_0$, we know that
$\Pr[B=t] \geq e^{-2e}\Pr[B'=t]$. Substituting this into the bound above
and using the fact that $\hat \eps = 3\eps$ and $\Pr[A'|_t \in S_t] = \Pr[A' \in S_t \,|\, B'=t]$, we get
\begin{eqnarray*}
\Pr[(A,B)\in T_1]   &\geq & \sum_{t \in Bad_1'} e^{\hat \eps}\Pr[A' \in S_t \,|\, B'=t]e^{-2\eps} \Pr[B'=t]  + \hat\delta\Pr[B \in Bad_1']  \\
&=& e^{\eps}\Pr[(A',B') \in T_1] + \hat\delta\Pr[B \in Bad_1'].
\end{eqnarray*}
By $(\eps,\delta)$-indistinguishability, $\Pr[(A,B)\in T_1] \leq
e^{\eps}\Pr[(A',B') \in T_1] + \delta$. Combining the upper and lower
bounds on the probability that $(A,B)\in T_1$, we have 
$\hat\delta\Pr[B \in Bad_1'] \leq \delta $, which implies that $$ \Pr[B \in Bad_1'] \leq \delta/\hat\delta\,.$$
By a similar argument, one gets that $\Pr[B \in Bad_2'] \leq \delta/\hat\delta.$ Finally,
\begin{eqnarray*}
  \Pr[B \in Bad_1 \cup Bad_2] 
  &\leq& \Pr[B \in Bad_0] +\Pr[B \in Bad_1']+\Pr[B \in Bad_2'] \\ 
  &=& \frac{2\delta}{1- e^{-\eps}}+\frac{\delta}{\hat\delta}+\frac{\delta}{\hat\delta}
      = \frac{2\delta}{1- e^{-\eps}}+\frac{2\delta}{\hat\delta} = \delta''.
\end{eqnarray*}

By symmetry, we also have $\Pr[B' \in Bad_1 \cup Bad_2] \leq
\frac{2\delta}{1- e^{-\eps}}+\frac{2\delta}{\hat\delta}$. Therefore,
with probability at least $1-\delta''$, $A|_{t}$ and $A'|_{t}$ are
$(\hat\eps,\hat\delta)$-indistinguishable, as claimed. 
\end{proof}


\subsection{The General Case: Proof of \thmref{ind2sdp}}
\label{sec:approxproof}

\newtheorem*{t2}{Theorem~\ref{thm:ind2sdp} (restated)}
\begin{t2}{~}\renewcommand{\labelenumi}{(\arabic{enumi})} 
\begin{enumerate}
\item \label{part:main1}
    If $\eps,\delta>0$ and $\delta< (1- e^{-\eps})^2/n$, then
    ($\eps,\delta$)-differential privacy implies
    $(\eps',\delta')$-semantic privacy on databases of size $n$ with
    $\eps'=e^{3\eps}-1+2\sqrt{n\delta}$ and $\delta' =
    4\sqrt{n\delta}$. 
\item\label{part:main2} If $\eps,\delta>0$ and $\eps \leq 0.45$, then
    $(\eps,\delta)$-semantic privacy implies
    $(3\eps,2\delta)$-differential privacy.
  \end{enumerate}
\end{t2}

\begin{proof}
(\ref{part:main1}) Let $\AAA$ be an $(\eps,\delta)$-differentially private algorithm. Let $b$ be any belief distribution and let $ \dx \sim b$.  Let $\AAA_i(\dx) =\AAA(\dx_{-i})$, i.e., $\AAA_i$ on input $\dx$ constructs $\dx_{-i}$ and then applies $\AAA$ on it. From \lemref{properties} (Part~\ref{it:ajoint}), we know that $(\dx,\AAA(\dx))$ and $(\dx,\AAA_i(\dx))$ are $(\eps,\delta)$-indistinguishable for every index $i=1,\dots,n$. 

Apply \lemref{bayes} with $\AAA(X) = \AAA(\dx)$,
$\AAA'(X)=\AAA_i(\dx)$, $\hat \eps = 3\eps$, and $\hat
\delta=\sqrt{n\delta}$. We get that with probability at least
$1-\delta''$ over $t\sim \AAA(\dx)$, the random variables
$\dx|_{\AAA(\dx)=t}$ and $\dx|_{\AAA_i(\dx)=t}$ are $(\hat
\eps,\hat\delta)$-indistinguishable, where $\delta '' \leq
2\delta/\hat \delta + 2\delta/(1- e^{-\eps}) \leq 4\sqrt{\delta/n}$. Note that $1- e^{-\eps}>\hat\delta=\sqrt{n\delta}$ (a condition assumed in the theorem).

Let $\delta' = n\delta''$; note that $\delta' \leq 4\sqrt{n\delta}$. Taking a union bound over all $n$ choices for the index $i$, we get that with probability at least $1-\delta'$ over the choice of $t\sim \AAA(\dx)$, all $n$ variables $\dx|_{\AAA_i(\dx)=t}$ (for different $i$'s) are $(\hat\eps,\hat\delta)$-indistinguishable from $\dx|_{\AAA(\dx)=t}$.

To complete the proof of (\ref{part:main1}), recall that  $(\hat\eps,\hat\delta)$-indistinguishability implies statistical distance at most $e^{3\hat \eps}-1+\hat \delta = \eps'$.
  
(\ref{part:main2})  To see that  $(\eps,\delta)$-semantic privacy
implies $(3\eps,2\delta)$-differential privacy, consider a belief
distribution $b$ which is uniform over two databases $\dx,\dy$ which
are at Hamming distance of one. The proof idea is same as in
Theorem~\ref{thm:eind}. Let $i$ be the position in which $\dx$ and
$\dy$ differ. Let $\A$ be an algorithm that satisfies
$(\eps,\delta)$-semantic privacy.

In Game $i$, 
$\dx$ and $\dy$ induce the same distribution on
transcripts, so the
distribution $\bar{b}_i[\cdot|t]$ will be uniform over $\dx$ and $\dy$
(for all transcripts $t$). 
We now turn to Game 0 (the real world). Let $E$ denote the set of
transcripts $t$ such that $\bar{b}_0[\cdot |t]$ assigns probabilities in $1/2\pm \eps$ to each
of the two databases $\dx$ and $\dy$.
Let $\bar\AAA$ denote the (random) output of $\A$ when run on a
database sampled from distribution $b$. 
The semantic privacy of $\A$  implies $E$ occurs with
probability at least $1-\delta$ over $t \sim \bar\AAA$.
Working through Bayes' rule as in
Theorem~\ref{thm:eind} shows that 
$$e^{-3\eps}\Pr[\AAA(\dy)=t] \leq \Pr[\AAA(\dx)=t] \leq e^{3\eps}
\Pr[\AAA(\dy)=t]$$ for all $t\in E$.
(This last step uses the assumption that $\eps
\leq 0.45$).
Moreover, since $\bar \AAA$ is an equal mixture of $\AAA(\dx)$ and $\AAA(\dy)$,
the event $E$ must occur with probability at least $1-2\delta$ under
both $t\sim \AAA(\dx)$ and $t\sim \AAA(\dy)$
Hence, $\AAA(\dx)$ and $\AAA(\dy)$ are
$(3\eps,2\delta)$-indistinguishable. Since this relationship holds for every pair of neighboring databases $\dx$ and $\dy$, means that $\AAA$ is $(3\eps,2\delta)$-differentially private.
\end{proof}

\section{Further Discussion}
\label{sec:discussion}
\thmref{ind2sdp} states that the relaxations of differential privacy
in some previous work still provide meaningful guarantees in the face
of arbitrary side information. This is {\em not} the case for all
possible relaxations, even very natural ones, as noted
in~\remref{statdiff}.

\paragraph{Calibrating Noise to a High-Probability Bound Local Sensitivity.}

In a different vein, the techniques used to prove \thmref{ind2sdp} can also be used to analyze schemes that do not provide privacy for {\em all} pairs of neighboring databases $\dx$ and $\dy$, but rather only for {\em most} such pairs (remember that neighboring databases are the ones that differ in one entry). Specifically, it is sufficient that those databases where the indistinguishability condition fails occur only with small probability. 

We first define a weakening of Definition~\ref{defn:epsdelta} so that it only holds for specific belief distributions.

\begin{definition}[($\eps,\delta$)-local semantic privacy]
A randomized algorithm is $(\eps,\delta)$-local semantically private for a belief distribution $b$ on $\mathcal{D}^n$ if  with probability at least $1-\delta$ over $t\sim\AAA(\dx)$ ($t$ drawn from $\AAA(\dx)$),  where the database $\dx$
is drawn according to $b$, and for all $i =1,\dots,n$:
$$\sd{\bar{b}_0[\cdot|t]\ ,\  \bar{b}_i[\cdot|t]\ } \leq \eps. $$
\end{definition}

\begin{theorem} \label{thm:dsemantic}
Let $\AAA$ be a randomized algorithm. Let $$\mathcal{E} = \{\dx : \forall \mbox{ neighbors }\dy \mbox{ of } \dx, \AAA(\dx) \mbox{ and } \AAA(\dy) \mbox{ are } (\eps,\delta)\mbox{-indistinguishable}\}.$$ Then $\AAA$ satisfies  $(\eps',\delta')$-local semantic privacy for any belief distribution $b$ such that $b[\mathcal{E}] = \Pr_{\dx \sim b}[\dx \in \mathcal{E}] \geq 1-\delta_1$ with $\eps'=e^{3\eps}-1+\sqrt{n\delta_2}$ and $\delta' \leq 4\sqrt{n\delta_2}$ as long as $\eps>\sqrt{n\delta_2}$, where $\delta_2 = \delta + \delta_1$.  
\end{theorem}
\begin{proof}
The proof is similar to \thmref{ind2sdp} (\ref{part:main1}).  Let $b$ be a belief distribution with $b[\mathcal{E}] \geq 1-\delta_1$ and let $\dx \sim b$. From \lemref{properties} (Part~\ref{it:joint}), we know that $(\dx,\AAA(\dx))$ and $(\dx,\AAA_i(\dx))$ are $(\eps,\delta+\delta_1)$-indistinguishable, where $\AAA_i(\dx)=\AAA(\dx_{-i})$. The remaining proof follows exactly as in \thmref{ind2sdp} (\ref{part:main1}). 
\end{proof}

We now discuss a simple consequence of the above theorem to the technique of adding noise according to {\em local sensitivity} of a function.
\begin{definition}[Local Sensitivity,~\cite{NRS07}]
For a function $f \,: \, \mathcal{D}^n \rightarrow \R$, and $\dx \in \mathcal{D}^n$, the local sensitivity of $f$ at $\dx$ is:
\[LS_f(\dx) = \max_{\dy \,:\, \dham(\dx,\dy)= 1} |f(\dx) - f(\dy)|.\]
\end{definition}

Let $Lap(\lambda)$ denote the Laplacian distribution. This distribution has density function $h(y) \propto \exp(-|y|/\lambda)$, mean $0$, and standard deviation $\lambda$. Using the Laplacian noise addition procedure of \cite{DMNS06,NRS07}, along with Theorem~\ref{thm:dsemantic} we get\footnote{Similar corollaries could be derived for other differential privacy mechanisms like those that add Gaussian noise instead of Laplacian noise.}, 
\begin{corollary}
Let $\mathcal{E} =\{ \dx \,:\, LS_f(\dx) \leq s \}$. Let $\AAA(\dx) = f(\dx) + \text{Lap}\left(\frac{s}{\eps}\right )$. Let $b$ be a belief distribution such that $b[\mathcal{E}] = \Pr_{\dx \sim b}[\dx \in \mathcal{E}] \geq 1-\delta_1$. Then $\AAA$ satisfies  $(\eps',\delta')$-local semantic privacy for belief distribution $b$ with $\eps'=e^{3\eps}-1+\sqrt{n\delta_1}$ and $\delta' \leq 4\sqrt{n\delta_1}$ as long as $\eps>\sqrt{n\delta_1}$. 
\end{corollary}
\begin{proof}
Let $\dx \sim b$. If $\dx \in \mathcal{E}$, then it follows from~\cite{DMNS06,NRS07}, that $\AAA(\dx)$ and $\AAA(\dx_{-i})$ are $(\eps,0)$-indistinguishable for every index $i = 1,\dots,n$. We can apply~\thmref{dsemantic} to complete the proof.
\end{proof}

The approach discussed here was generalized significantly by~\citet{BassilyGKS13}; we refer to their work for a detailed discussion.


\section*{Acknowledgements} 
We are grateful for helpful discussions with Cynthia Dwork, Daniel
Kifer, Ashwin Machanvajjhala, Frank McSherry, Moni Naor, Kobbi Nissim,
and Sofya Raskhodnikova. 

\addcontentsline{toc}{section}{References}
\bibliographystyle{plainnat}
\bibliography{master}

\appendix \section*{Appendix}
\section{Proof of \lemref{properties}}
\label{app:propertiesproof}

\noindent{\textit{Proof of Part \ref{it:pw2ind}.}} 
Let $Bad$ be the set of {\em bad} values of $a$, that is $$Bad = \{a \,:\, \Pr[X=a] < e^{-\eps} \Pr[Y=a] \mbox{ or }  \Pr[X=a]> e^{\eps} \Pr[Y=a] \}.$$ By definition, $\Pr[X\in Bad]\leq \delta$. Now
consider any set $S$ of outcomes. $$\Pr[X\in S] \leq \Pr[X \in S \setminus Bad] + \Pr[X\in Bad].$$ The first term is at most $e^\eps \Pr[Y\in S \setminus Bad]\leq e^\eps \Pr[Y\in S]$. Hence, $\Pr[X \in S] \leq e^\eps \Pr[Y\in S] + \delta$, as required. The case of $\Pr[Y\in S]$ is symmetric. Therefore, $X$ and $Y$ are $(\eps,\delta)$-indistinguishable. \\

\noindent{\textit{Proof of Part \ref{it:ind2pw}.}} Let $S= \{a \,:\, \Pr[X=a] > e^{2\eps} \Pr[Y=a]\}$. Then
\[  \Pr[X \in S] > e^{2\eps} \Pr[Y \in S]\,. \]
By
$(\epsilon,\delta)$ indistinguishability,  we have
$
\delta \geq \Pr[X \in S] - e^{\eps}\Pr[Y \in S] > (e^{2\eps}-
e^{\eps}) \Pr[Y \in S]$. Equivalently,
\begin{equation}
 \Pr[Y \in S] <
\frac{\delta}{e^{2\eps} - e^\eps} = \frac{\delta}{e^{2\eps} (1-e^{-\eps})}\, .\label{eq:5}
\end{equation}
Now consider the set $S'=\{a \,:\, \Pr[X=a] < e^{-2\eps}
\Pr[Y=a]\}$. A symmetric argument to the one above shows that
$\Pr[X \in S'] < \delta/(e^{2\eps}-e^{\eps}) $. Again using
indistinguishability, we get 

\begin{equation}
\Pr[Y\in S'] \leq e^{\eps} \Pr[X \in S'] + \delta < e^{\eps}
\frac{\delta}{e^{2\eps} - e^\eps} + \delta = \frac{\delta
  }{1-e^{-\eps}}\, . \label{eq:4}
\end{equation}

The bound of \eqref{eq:4} is always larger than that of \eqref{eq:5},
so we have  $\Pr[Y \in S \cup S'] \leq \delta \cdot \frac{2
  e^{\eps}}{e^{\eps}-1}$. We can get the same bound on $\Pr[X \in S
\cup S']$ by symmetry.
Therefore, with
probability at least $1-\delta \cdot \frac{2
  e^{\eps}}{e^{\eps}-1}$ for $a$ drawn from the distribution of
either $X$ or $Y$ we have:
$e^{-2\eps}\Pr[Y=a] \leq \Pr[X=a] \leq e^{2\eps} \Pr[Y=a]$. \\

\noindent{\textit{Proof of Part \ref{it:ajoint}.}} Let $(X,\A(X))$ and $(X,\A'(X))$ be random variables on $D \times E$. Let $S$ be an arbitrary subset of $D \times E$ and, for every $a \in D$, define $S_a =\{b \in E\,:\, (a,b) \in S\}$. 
\begin{eqnarray*}
\Pr[(X,\A(X)) \in S] &\leq& \sum_{a \in D}\Pr[\A(X) \in S_a \,|\, X=a]\Pr[X=a] \\
& \leq & \sum_{a \in D}(e^{\eps}\Pr[\A'(X) \in S_a \,|\, X=a]+\delta)\Pr[X=a] \\
& \leq & \delta + e^{\eps}\Pr[(X,\A'(X)) \in S].
\end{eqnarray*}
By symmetry, we also have $\Pr[(X,\A'(X)) \in S] < \delta+e^{\eps}\Pr[(X,\A(X)) \in S]$. Since the above inequalities hold for every selection of $S$, implies that $(X,\A(X))$ and $(X,\A'(X))$ are $(\eps,\delta)$-indistinguishable. \\

\noindent{\textit{Proof of Part \ref{it:joint}.}}  Let $(X,\A(X))$ and $(X,\A'(X))$ be random variables on $D \times E$. Let $T \subset D$ be the set of $a$'s for which $\A(a) \leq e^{\eps} \A'(a)$. Now, let $S$ be an arbitrary subset of $D \times E$ and, for every $a \in D$, define $S_a =\{b \in E\,:\, (a,b) \in S\}$. 
\begin{eqnarray*}
\Pr[(X,\A(X)) \in S] &=&  \sum_{a \notin T}\Pr[\A(X) \in S_a \,|\, X=a]\Pr[X=a]  +  \sum_{a \in T}\Pr[\A(X) \in S_a \,|\, X=a]\Pr[X=a] \\
& \leq & \sum_{a \notin T} \Pr[X=a]  +  \sum_{a \in T}\Pr[\A(X) \in S_a \,|\, X=a]\Pr[X=a] \\
&=& \Pr[X \notin T] + \sum_{a \in T}\Pr[\A(X) \in S_a \,|\, X=a]\Pr[X=a] \\
& \leq & \delta_1 + \sum_{a \in T}(e^{\eps}\Pr[\A'(X) \in S_a \,|\, X=a]+\delta)\Pr[X=a] \\
& \leq & \delta + \delta_1 + e^{\eps}\Pr[(X,\A'(X)) \in S].
\end{eqnarray*}
By symmetry, we also have $\Pr[(X,\A'(X)) \in S] < \delta+\delta_1 + e^{\eps}\Pr[(X,\A(X)) \in S]$. Since the above inequalities hold for every selection of $S$, implies that $(X,\A(X))$ and $(X,\A'(X))$ are $(\eps,\delta + \delta_1)$-indistinguishable. \\


\noindent{\textit{Proof of Part \ref{it:sd}.}} Let $X$ and $Y$ be random variables on $D$. By definition $\sd{X,Y}=\max_{S \subset D}|\Pr[X \in S] - \Pr[Y \in S]|$. 
For any set $S \subset D$, 
\begin{eqnarray*}
\lefteqn{2|\Pr[X  \in S]-\Pr[Y \in S]|} \\
&=& \left |\Pr[X  \in S]-\Pr[Y  \in S]\right | + \left |\Pr[X  \notin S]-\Pr[Y  \notin S] \right | \\
&=& \left |\sum_{c \in S}(\Pr[X =c]-\Pr[Y =c])\right | + \left |\sum_{c \notin S}(\Pr[X =c]-\Pr[Y =c])\right | \\
&\leq& \sum_{c \in S}\left |\Pr[X =c]-\Pr[Y =c] \right | + \sum_{c \notin S}\left |\Pr[X =c]-\Pr[Y =c]\right |  \\
&=& \sum_{c \in D } \left |\Pr[X =c]-\Pr[Y =c] \right | \\ &\leq& \sum_{c \in D}(e^{\eps}\Pr[Y =c] + \delta - \Pr[Y =c]) + \sum_{c \in D}(e^{\eps}\Pr[X =c] +\delta - \Pr[X =c]) \\
&=& 2\delta + (e^{\eps}-1) \sum_{c \in D}\Pr[Y =c] + (e^{\eps}-1) \sum_{c \in D}\Pr[X =c] \\ &\leq & 2(e^{\eps}-1)+2\delta=2\bar{\eps}+ 2\delta.
\end{eqnarray*}
This implies that $|\Pr[X  \in S]-\Pr[Y \in S]| \leq \bar\eps + \delta$. Since the above inequality holds for every $S \subset D$, it immediately follows that the statistical difference between $X$ and $Y$ is at most $\bar{\eps}+\delta$. \qed

\section{Another View of Semantic Privacy}\label{app:anotherview}
\newtheorem*{thma}{Theorem A.3} 
\newtheorem*{clma}{Theorem A.2}
\newtheorem*{defna}{Definition A.1}

In this section, we discuss another possible definition of $(\eps,\delta)$-semantic privacy. Even though this definition seems to be the more desirable one, it also seems hard to achieve. 
\begin{defna} [reality-oblivious ($\eps,\delta$)-semantic privacy] 
A randomized algorithm is reality-oblivious $(\eps,\delta)$-semantically private if for all belief distributions $b$ on $\mathcal{D}^n$, for all databases $\dx\in \mathcal{D}^n$, with probability at least $1-\delta$ over transcripts $t$ drawn from $\A(\dx)$, and for all $i=1,\dots,n$:
$$\sd{\bar{b}_0[\cdot|t]\ ,\  \bar{b}_i[\cdot|t]\ } \leq \eps. $$
\end{defna}

We prove that if the adversary has arbitrary beliefs, then
$(\eps,\delta)$-differential privacy doesn't provide any reasonable
reality-oblivious $(\eps',\delta')$-semantic privacy guarantee. 

\begin{clma}\label{thm:noind2sdp}
($\eps,\delta$)-differential privacy does not imply reality-oblivious $(\eps',\delta')$-semantic privacy for any reasonable values of $\eps'$ and $\delta'$. 
\end{clma}
\begin{proof} This counterexample is due to Dwork and McSherry: suppose that the belief distribution is uniform over $\{(0^n),(1,0^{n-1})\}$, but that real database is $(1^{n})$. Let the database $\dx=(x_1,\dots,x_n)$. Say we want to reveal $f(\dx)=\sum_ix_i$. Adding Gaussian noise with variance $\sigma^2=\logdel/\eps^2$ satisfies $(\eps,\delta)$-differential privacy (refer \cite{DMNS06,NRS07} for details). However, with overwhelming probability the output will be close to $n$, and this will in turn induce a very non-uniform distribution over $\{(0^n),(1,0^{n-1})\}$ since $(1,0^{n-1})$ is exponentially (in $n$) more likely to generate a value near $n$ than $(0^n)$. More precisely, due to the Gaussian noise added,
$$\frac{\Pr[\AAA(\dx)=n \, | \, \dx=(0^n)]}{\Pr[\AAA(\dx)= n \, | \, \dx=(1,0^{n-1})]} = \frac{\exp\left (\frac{-n^2}{2\sigma} \right)}{\exp \left(\frac{-(n-1)^2}{2\sigma}\right )} = \exp\left (\frac{-2n+1}{2\sigma} \right).$$
Therefore, given that the output is close to $n$, the posterior distribution of the adversary would be exponentially more biased toward $(1,0^{n-1})$ than $(0^n)$. Hence, it is exponentially far away from the prior distribution which was uniform. On the other hand, on $\dx_{-1}$,  no update will occur and the posterior distribution will remain uniform over $\{(0^n),(1,0^{n-1})\}$ (same as the prior). Since the posterior distributions in these two situations are exponentially far apart (one exponentially far from uniform, other uniform), it shows that ($\eps,\delta$)-differential privacy does not imply any reasonable guarantee on reality-oblivious semantic privacy.
\end{proof}

The counterexample of \thmref{noind2sdp} implies that adversaries
whose belief distribution is very different from the real database may
observe a large change in their posterior distributions. We do not
consider this a `violation of ``privacy'', since the issue lies in the
incorrect beliefs, not the mechanism per se.

\end{document}